\documentclass[letterpaper, 10 pt, conference]{ieeeconf} 
\IEEEoverridecommandlockouts                            
\makeatletter

\let\proof\@undefined
\let\endproof\@undefined
\let\labelindent\@undefined 
\makeatother

\usepackage[dvipsnames]{xcolor}
\usepackage{todonotes}
\usepackage{enumitem}
\usepackage{graphicx,float}
\usepackage{siunitx}
\usepackage{booktabs}
\usepackage{amsmath,amsfonts,amssymb,amscd} 
\usepackage{mathtools,mathrsfs,bm,amsthm}
\newcommand{\colvec}[2][.9]{%
  \scalebox{#1}{%
    \renewcommand{\arraystretch}{1}%
    $\begin{bmatrix}#2\end{bmatrix}$%
  }
}
\usepackage{caption}
\usepackage{subcaption}

\usepackage{standalone}
\usepackage{soul}

\usepackage{cancel}

\usepackage{hyperref}
\hypersetup{%
   colorlinks=true,%
   citecolor = bondiblue,%
   urlcolor  = bondiblue,%
   linkcolor = MidnightBlue%
}

\newtheorem{remark}{Remark}
\newtheorem{problem}{Problem}

\usepackage[
placement=top,
angle=0,
color=red,
scale=1.1,
hshift=0,
vshift=-15
]{background}
\backgroundsetup{contents=\bf{----------------------------------------------------------------- PREPRINT -----------------------------------------------------------------}}
\title{\LARGE \bf
Trajectory Tracking for Tilted Hexarotors\\ with Concurrent Attitude Regulation% on $SO(3)$
}

\author{Marco Perin$^{1}$, Massimiliano Bertoni$^{2}$, Giulia Michieletto$^{2}$, Roberto Oboe$^{2}$, and Angelo Cenedese$^{1}$% <-this % stops a space
\thanks{*{This work is partially supported by the EU Next-Generation (PNRR) within the Italian National Ph.D. Program in Autonomous Systems (DAuSy), and by MUR through PRIN Grant DOCEAT 2020RTWES4.}}% <-this % stops a space
\thanks{$^{1}$ M.Perin and A.Cenedese are with the Department of Information Engineering, University of Padova, Italy.
        }%
\thanks{$^{2}$M.Bertoni, G.Michieletto, and R.Oboe are with the Department of Management and Engineering, University of Padova, Italy.
        }%
\thanks{\hspace{-9pt}{Contacts: M.Perin - 
{\tt\small marco.perin.6@studenti.unipd.it}}}
\thanks{\hspace{+24pt}M.Bertoni - {\tt\small massimiliano.bertoni@phd.unipd.it}}%
}

\begin{document}

\maketitle
\thispagestyle{empty}
\pagestyle{empty}

%%%%%%%%%%%%%%%%%%%%%%%%%%%%%%%%%%%%%%%%%%%%%%%%%%%%%%%%%%%%%%%%%%%%%%%%%%%%%%%%
\begin{abstract}
Tilted hexarotors embody a technology that remains partially unexploited in terms of its potential, especially concerning precise and concurrent position and attitude control.
Focusing on these aerial platforms, we propose two control architectures that can tackle the trajectory tracking task, ensuring also the attitude regulation: one is designed resting on the differential flatness property of the system, which is investigated in the paper, and the other is a hierarchical nonlinear controller. 
We comparatively discuss the performance of the two control schemes, in terms of the accuracy of both the tracking control action and the attitude regulation, the input effort, and the robustness in the presence of disturbances. 
Numerical results reveal both the robustness of the hierarchical approach in the case of external disturbance and the accuracy of the differential flatness-based controller in unwindy conditions. 
\end{abstract}

%%%%%%%%%%%%%%%%%%%%%%%%%%%%%%%%%%%%%%%%%%%%%%%%%%%%%%%%%%%%%%%%%%%%%%%%%%%%%%%%
\section{INTRODUCTION}

Over the last decade, Unmanned Aerial Vehicles (UAVs) have seen a growing interest in robotic research motivated by the emerging challenges they pose in the design of estimation and control solutions and the versatility they show in a vast application domain. Aerial platforms represent a key technology in many fields, ranging from traditional monitoring operations to cutting-edge physical interaction tasks within rural, civil, and industrial contexts~\cite {elmeseiry2021detailed}. 
The interest in developing efficient and robust solutions for modern applications has led to the design of new UAV configurations with improved actuation capabilities. 
In this sense, the study of fully-actuated aerial platforms has joined that of the under-actuated coplanar and collinear quadrotors, and the literature devoted to multi-rotor UAVs having more than four, even tilted or tilting, propellers is recently boosting~\cite{rashad2020fully}. 

Among the state-of-the-art UAVs, the star-shaped hexarotor having tilted propellers evenly spaced on a circumference (hereafter referred to as Tilted HexaRotor - \textit{TedHR})  has proved to be the configuration with the minimum number of rotors guaranteeing both the full actuation and the robustness to the failure of any propeller~\cite{michieletto2018fundamental}.
In particular, the control force and the control moment of this class of aerial platforms can be regulated in a completely independent manner and can be assigned in a region of the 3D space proportionally depending on the propellers' tilt angles, although at the cost of spurious components in static hovering conditions
~\cite{tadokoro2017maneuverability}. %,ryll2021fast}.
\smallskip

\noindent{\textit{Related works -}} When in static hovering, a UAV is required to fly in a desired position with null linear and angular velocities, often while also maintaining a desired orientation. The most popular control approaches designed to keep a TedHR in this flight condition involve cascaded architectures based on geometric paradigm~\cite{michieletto2017control}, nonlinear strategies~\cite{michieletto2017nonlinear,michieletto2020hierarchical}, port-Hamiltonian approaches~\cite{rashad2019port}. 
Extensive literature with successful simulations and experimental tests is also available on the path following task for TedHRs. Many works propose robust solutions in case of uncertainties and disturbances exploiting feedback linearization strategies~\cite{antonello2018dual}, adaptive techniques~\cite{arizaga2019adaptive}, robust nonlinear methods~\cite{flores2022robust}. 
In most cases, the attention is limited to position tracking without accounting for any attitude reference. Guaranteeing the attitude regulation while following a reference position profile is, indeed, a less popular problem although its full actuation allows the TedHR to track both position and attitude references at the same time. 
Typical control solutions entail the computation of the wrench required to compensate for the nonlinear dynamical effects and to zero the pose (position and attitude) tracking error~\cite{rajappa2015modeling}, though more sophisticated full-pose controllers have been recently proposed in~\cite{franchi2018full} and~\cite{hamandi2023full}. 
In the former case, a geometric approach is adopted dealing with $SE(3)$. In the latter one, the pose trajectory tracking problem is tackled in an optimization framework guaranteeing the online computation of feasible control inputs while modifying the reference attitude to satisfy the actuation constraints. 
\smallskip

\noindent{\textit{Contributions -}}
Focusing on the class of star-shaped tilted hexarotors, we study their differential flatness properties and then we propose two different control architectures to tackle the position trajectory tracking task ensuring also the attitude regulation along the three inertial axes in $SO(3)$. 
The designed controllers are one based on the differential flatness of the TedHRs (\textit{Flatness-based Controller - FC)}, which represents also a reference benchmark, and one characterized by a hierarchical architecture (\textit{Hierarchical Controller - HC)}.
As for the former, while a similar approach is already present in the literature, to the best of the authors' knowledge only collinear UAVs have been considered, specifically referring to quadrotors. The state-of-art works, also, do not take into account the full pose as a flat output but employ only the position for this aim, relying on internal attitude controllers to accommodate the platform orientation.
In the few cases where differential flatness is applied to hexarotor UAVs, collinear platforms are still involved, hence under-actuated vehicles~\cite{michieletto2018fundamental}. This fact leads to the possibility of achieving solely position and yaw control. 
To address this issue, in our work, we exploit the full actuation of the TedHRs to formalize the differential flatness problem and provide a control scheme based on a full pose trajectory flat output.
The other implemented controller is an improved version of the hierarchical nonlinear control architecture introduced in~\cite{michieletto2017nonlinear} and then refined in~\cite{michieletto2020hierarchical}. More specifically, we generalize the previous control solution to cope with the position trajectory tracking along with attitude regulation tasks, instead of limiting to static hovering regulation as in~\cite{michieletto2020hierarchical}.
To this aim, the regulator structure is revisited and adapted to incorporate specific feedforward terms provided by the trajectory planner. This adjustment makes the closed-loop system able to attain zero 3D tracking error in position and zero steady-state error when constant attitude references are imposed.

The controllers' performances are evaluated in the MATLAB-Simulink environment, both in ideal conditions and in more realistic scenarios obtained by employing wind models that include wind gusts, Dryden, and shear models.
These numerical simulations provide good insights into the robustness of the two control architectures and their ability to possibly withstand even high disturbances.
\smallskip 

\noindent{\textit{Paper Organization -}} The rest of the paper is organized as follows. Section~\ref{sec:scenario} is devoted to the formalization of the trajectory tracking with attitude regulation problem.  Section~\ref{sec:flatness} describes the designed FC solution, while the HC architecture is outlined in Section~\ref{sec:hierarchical}. Section~\ref{sec:validation} provides the results of the numerical validation of the two control solutions. The main conclusions and future research directions are summarized in Section~\ref{sec:conclusions}.

%%%%%%%%%%%%%%%%%%%%%%%%%%%%%%%%%%%%%%%%%%%%%%%%%%%%%%%%%%%%%%%%%%%%%%%%%%%%%%%%
\section{TRAJECTORY TRACKING WITH\\ ATTITUDE REGULATION TASK}
\label{sec:scenario}

In this work, we focus on the class of tilted hexarotor platforms having a star-shaped configuration. These multi-rotor UAVs are actuated by six propellers evenly spaced on a circumference centered in the vehicle center of mass (CoM) and spinning about tilted axes in alternate directions. Specifically, we consider the case wherein the rotor spinning axes are tilted both along the direction identified by the vehicle arm and along the orthogonal vertical one,
and the corresponding tilt angles are fixed during flight. Formally, introducing the reference frame $\mathscr{F}_B=\{O_B, (\mathbf{x}_B, \mathbf{y}_B, \mathbf{z}_B)\}$ centered in the vehicle CoM (\textit{body frame)}, the direction of any $i$-th rotor spinning axis $\mathbf{z}_{P_i} \in \mathbb{R}^3$, $i \in \{1 \ldots 6\}$ is time-invariant in $\mathscr{F}_B$ and such that $\mathbf{z}_{P_i}=\mathbf{R}_x((-1)^{i}\alpha)\mathbf{R}_y(\beta) \mathbf{z}_B$ with $\mathbf{R}_x(\cdot), \mathbf{R}_y(\cdot) \in SO(3)$ denoting the elementary rotation around $x,y$-axis of the given angle $\alpha, \beta \in [-\pi, \pi)$. 

Such structural features ensure the full actuation and the decoupling of the transitional and rotational dynamics of these TedHR platforms which can be modeled as rigid bodies in 3D space. Formally, accounting for $\mathscr{F}_B$, the pose of a TedHR with respect to the inertial reference frame $\mathscr{F}_W=\{O_W, (\mathbf{x}_W, \mathbf{y}_W, \mathbf{z}_W)\}$ (\textit{world frame)} is identified by the vector $\mathbf{p} \in \mathbb{R}^3$, defining the position of $O_B$ in $\mathscr{F}_W$, and the unit quaternion $\mathbf{q} = \colvec{\eta \;\; \boldsymbol{\epsilon}^\top}^\top \in \mathbb{S}^3$, representing the relative orientation between $\mathscr{F}_B$ and $\mathscr{F}_W$. 

\begin{remark}
As far as the TedHR's orientation is concerned, we mainly use the quaternion representation and for any $\mathbf{q} \in \mathbb{S}^3$ we indicate with $\eta \in \mathbb{R}$ and $\boldsymbol{\epsilon} \in \mathbb{R}^3$ its scalar and vector part, respectively. 
Nonetheless, we also resort to the rotation matrices and the Euler angles representations. In datil, we denote with $\mathbf{R}(\mathbf{q})$ the rotation matrix in $SO(3)$ associated to $\mathbf{q}$, and we assume that any matrix $\mathbf{R}(\mathbf{q})$ is a function of the Euler angles $\boldsymbol{\delta}=\colvec{ \phi \; \theta \; \psi}^\top$ according to the rotation composition based on the ZYX sequence.
\end{remark}

Then, considering the platform linear velocity $\mathbf{v} \in \mathbb{R}^3$ expressed in world frame and its angular velocity $\boldsymbol{\omega} \in \mathbb{R}^3$ expressed in body frame, the kinematics and dynamics of the TedHR result to be governed by the following equations 
\begin{subequations}\label{eq:dyn}
\begin{align}
\dot{\mathbf{p}}&= \mathbf{v} \\
 \dot{\mathbf{q}}&=\frac{1}{2} \mathbf{q} \circ \colvec{0 \\ \boldsymbol{\omega} }= \frac{1}{2} \colvec{
 -\boldsymbol{\epsilon}^\top \\ \eta\mathbf{I}_3 - 
 \left[ \boldsymbol{\epsilon}\right]_\times
 } \boldsymbol{\omega} \label{eq:kin_rot}\\
m \ddot{\mathbf{p}}&=-m g \mathbf{e}_{3}+\mathbf{R}(\mathbf{q}) \mathbf{F} \mathbf{u} \label{eq:dyn_trasl} \\
\mathbf{J} \dot{\boldsymbol{\omega}}&=-\boldsymbol{\omega} \times \mathbf{J} \boldsymbol{\omega}+\mathbf{M u} \label{eq:dyn_rot}
\end{align}
\end{subequations}
where $\circ$ indicates the quaternion composition operation, $m, g>0$ denote the UAV mass and the gravitational constant respectively, $\mathbf{J} \in \mathbb{R}^{3 \times 3}$ represents the vehicle inertia matrix in $\mathscr{F}_{B}$,  $ \left[ \boldsymbol{\epsilon}\right]_\times$ stands for the skew-symmetric matrix associated to the vector $\boldsymbol{\epsilon}$, and $\mathbf{e}_{3} \in \mathbb{R}^{3}$ refers to the third column of the identity matrix  $\mathbf{I}_{3} \in \mathbb{R}^{3 \times 3}$, identifying the direction of $\mathbf{z}_W$. 
In~\eqref{eq:dyn_trasl}-\eqref{eq:dyn_rot} the vector $\mathbf{u} \in \mathbb{R}^6$ constitutes the TedHR command input, stacking the assignable squared propellers spinning rates. Thus, the matrices $\mathbf{F}, \mathbf{M} \in \mathbb{R}^{3 \times 6}$ respectively represent the control force and moment input matrices and depend on the geometric and aerodynamic characteristics of the platform. In detail, we have that
\begin{align}
\label{eq:control_force_moment}
    \mathbf{f}_c = \mathbf{F} \mathbf{u} \quad \text{and} \quad \boldsymbol{\tau}_c = \mathbf{M} \mathbf{u}
\end{align}
with $\mathbf{f}_c, \boldsymbol{\tau}_c \in \mathbb{R}^3$ denoting the control force and the control moment expressed in the body frame. 

\begin{remark}
    The control input matrices can be interpreted as a function of the tilt angles $\alpha$ and $\beta$. Specifically, the condition $(\alpha, \beta) \neq (0,0)$ guarantees the full rank property for both $\mathbf{F}$ and $\mathbf{M}$,  as for the matrix $\colvec{\mathbf{F}^\top \; \mathbf{M}^\top}^\top \in \mathbb{R}^{6 \times 6}$. This last fact ensures the full actuation of the UAV.
\end{remark}

For the described star-shaped tilted hexarotor platforms, we address the following control problem. 

\begin{problem}[Trajectory Tracking with concurrent Attitude Regulation]
Design a control solution that guarantees zero tracking error for a dynamic reference position $\mathbf{p}_r \in \mathbb{R}^3$ and concurrently for a piecewise constant attitude reference $\mathbf{q}_r \in \mathbb{S}^3$ (or equivalently $\boldsymbol{\delta}_r \in (\mathbb{S}^1)^3$ 
 so that $\mathbf{R}(\mathbf{q}_r )= \mathbf{R}(\boldsymbol{\delta}_r)$).
\label{prb:almost6Dcontrol}
\end{problem}

%%%%%%%%%%%%%%%%%%%%%%%%%%%%%%%%%%%%%%%%%%%%%%%%%%%%%%%%%%%%%%%%%%%%%%%%%%%%%%%%

\section{FLATNESS-BASED CONTROLLER}
\label{sec:flatness}

In the following, we first prove that any TedHR turns out to be a differentially flat system given a suitable choice of the state, input, and output vectors (Section~\ref{sec:diff_flatness}). Then, we describe the FC structure which exploits such a property and is characterized by an ad-hoc feedback action (Section~\ref{sec:FC_structure}). 

\subsection{TedHR Differential Flatness}
\label{sec:diff_flatness}

A system is said to be \textit{differentially flat}
if it is possible to express its states and inputs
as functions of a set of outputs and a finite number of its derivatives (\textit{flat outputs}). 
Formally, introducing the state, input, and output vectors,
namely $\mathbf{x} \in \mathbb{R}^n$, $\boldsymbol{\mu} \in \mathbb{R}^m$ and $\mathbf{y} \in \mathbb{R}^m$ with $n,m \in \mathbb{N}$, for a differentially flat system it is possible to identify the functions $g_x(\cdot)$ and $g_\mu(\cdot)$ such that
\begin{equation}
\label{eqn:diff_flat:flatness_props}
    \mathbf{x}     =  g_x(\mathbf{y}, \dot{\mathbf{y}},\ddot{\mathbf{y}}, \dots) \; \text{and} \; 
    \boldsymbol{\mu}  =  g_\mu(\mathbf{y}, \dot{\mathbf{y}},\ddot{\mathbf{y}}, \dots).
\end{equation}

To assess the differential flatness of the TedHR platforms, we select the state vector as $\mathbf{x} = \colvec{ \mathbf{p}^\top \; \mathbf{v}^\top \; \boldsymbol{\delta}^\top \; \boldsymbol{\omega}^\top}^\top \in \mathbb{R}^{12}$, so that the eqs.~\eqref{eq:dyn} lead to the linear state-space system
\begin{equation}
   \label{eqn:model:ss:full}
   \dot{\mathbf{x}}  = \mathbf{A}(\mathbf{x}) \mathbf{x} + \mathbf{B} \boldsymbol{\mu} - \mathbf{g}
\end{equation}
where $\mathbf{g}  \in \mathbb{R}^{12}$ stands for the gravity vector, i.e,  $ \mathbf{g} = g \colvec{
      \mathbf{0}^\top   \;   \mathbf{e}_3^\top \;
      \mathbf{0}^\top   \;
      \mathbf{0}^\top}^\top$ being $\mathbf{0} \in \mathbb{R}^3$ the (column) zero vector. 
The matrices $\mathbf{A}(\mathbf{x}) \in \mathbb{R}^{12 \times 12} $ and $\mathbf{B} \in \mathbb{R}^{12 \times 6}$ in~\eqref{eqn:model:ss:full} are defined as 
\begin{equation}
   \mathbf{A}(\mathbf{x})=\colvec{
      \mathbf{0}_{3} & \mathbf{I}_{3} & \mathbf{0}_{3} & \mathbf{0}_{3}                          \\
      \mathbf{0}_{3} & \mathbf{0}_{3} & \mathbf{0}_{3} & \mathbf{0}_{3}                          \\
      \mathbf{0}_{3} & \mathbf{0}_{3} & \mathbf{0}_{3} & \mathbf{W}(\boldsymbol{\delta})^{-1} \\
      \mathbf{0}_{3} & \mathbf{0}_{3} & \mathbf{0}_{3} & \mathbf{0}_{3}              } \quad  \mathbf{B} = \colvec{
      \mathbf{0}_{3}   & \mathbf{0}_{3}       \\
      \frac{1}{m}\mathbf{I}_{3} & \mathbf{0}_{3}       \\
      \mathbf{0}_{3}   & \mathbf{0}_{3}       \\
      \mathbf{0}_{3}   & \mathbf{J}^{-1} }
\end{equation}
with $\mathbf{0}_3 \in \mathbb{R}^{3 \times 3}$ denoting the zero matrix and $\mathbf{W}(\boldsymbol{\delta}) \in \mathbb{R}^{3 \times 3}$ depending on the UAV attitude represented thought the Euler angles convention so that
\begin{equation}
     \boldsymbol{\omega} = \mathbf{W}(\boldsymbol{\delta})\dot{\boldsymbol{\delta}}, \quad \mathbf{W}(\boldsymbol{\delta}) = \colvec{ 1 & 0 &-s\theta\\ 0 & c\phi & c\theta s\phi \\ 0 & -s\phi & c\theta c\phi} 
\end{equation}
where we use the notation $c\cdot$ and $s\cdot$ to indicate the cosine and sine functions. 
The input vector $\boldsymbol{\mu}$ in~\eqref{eqn:model:ss:full} (\textit{flat input}) is related to the input vector $\mathbf{u}$ in~\eqref{eq:dyn} (\textit{dynamics input}). In particular, it holds that
\begin{equation}
\label{eqn:model:ss:input}
   \boldsymbol{\mu} 
     = f(\mathbf{x}, \mathbf{u})
   =
  \colvec{\mathbf{0}_3 \\  -\boldsymbol{\omega}\times \mathbf{J} \boldsymbol{\omega}} + \colvec{
      \mathbf{R}(\boldsymbol{\delta}) & \mathbf{0}_3 \\ \mathbf{0}_3 & \mathbf{I}_3} \colvec{ \mathbf{F} \\
       \mathbf{M}
}\mathbf{u}.
\end{equation}
Finally, we define the (flat) output vector by accounting for the position and the orientation of the TedHR, namely, we select $\mathbf{y}=\colvec{\mathbf{p}^\top \; \boldsymbol{\delta}^\top}^\top \in \mathbb{R}^3 \times (\mathbb{S}^1)^3$. With this choice, we have that
\begin{equation}
\label{eqn:model:ss:output}
   \mathbf{y} = h(\mathbf{x})
   = \mathbf{C}\mathbf{x}
   =
   \colvec{
      \mathbf{I}_3 & \mathbf{0}_3 & \mathbf{0}_3 & \mathbf{0}_3 \\
      \mathbf{0}_3 & \mathbf{0}_3 & \mathbf{I}_3 & \mathbf{0}_3 \\
   }\mathbf{x}.
\end{equation}

Exploiting~\eqref{eqn:model:ss:full}, \eqref{eqn:model:ss:input}, and~\eqref{eqn:model:ss:output}, one can verify that the TedHR is a differentially flat system since it holds that
\begin{equation}
\label{eq:diff_state}
      \mathbf{x} = g_x(\mathbf{y}, \dot{\mathbf{y}}) =
      \colvec{
         \mathbf{I}_3 & \mathbf{0}_3 \\
         \mathbf{0}_3 & \mathbf{0}_3 \\
         \mathbf{0}_3 & \mathbf{I}_3 \\
         \mathbf{0}_3 & \mathbf{0}_3 \\
     }    \mathbf{y}
      +
      \colvec{
         \mathbf{0}_3 & \mathbf{0}_3              \\
         \mathbf{I}_3 & \mathbf{0}_3              \\
         \mathbf{0}_3 & \mathbf{0}_3              \\
         \mathbf{0}_3 & \mathbf{W}(\boldsymbol{\delta})^{-1} \\
      } \dot{\mathbf{y}},
\end{equation}
and $\boldsymbol{\mu}= g_{\mu}(\mathbf{y}, \dot{\mathbf{y}}, \ddot{\mathbf{y}})$ because of~\eqref{eqn:model:ss:input} and given that  the input vector $\mathbf{u}$ can in turn be expressed as $\mathbf{u} = g_u (\mathbf{y}, \dot{\mathbf{y}}, \ddot{\mathbf{y}})$, i.e.,
\begin{equation}
\label{eq:diff_input}
          \mathbf{u} =  g_u (\mathbf{y}, \dot{\mathbf{y}}, \ddot{\mathbf{y}}) 
          = \colvec{ \mathbf{F} \\ \mathbf{M} }^{-1}
      \colvec{
         m\mathbf{R}(\boldsymbol{\delta})^{\top}(\ddot{\mathbf{p}} + g \mathbf{e}_3) \\
         \mathbf{J} \dot{\boldsymbol{\omega}} + \boldsymbol{\omega} \times \mathbf{J} \boldsymbol{\omega}
      }.
\end{equation}
Thus, it results  $\boldsymbol{\mu} = g_{\mu}(\mathbf{y}, \dot{\mathbf{y}}, \ddot{\mathbf{y}})= f(g_x(\mathbf{y}, \dot{\mathbf{y}}), g_u (\mathbf{y}, \dot{\mathbf{y}}, \ddot{\mathbf{y}}))$.

\subsection{Controller Architecture}
\label{sec:FC_structure}

\begin{figure}[t!]
   \centering
   \includegraphics[width=1.0\columnwidth]{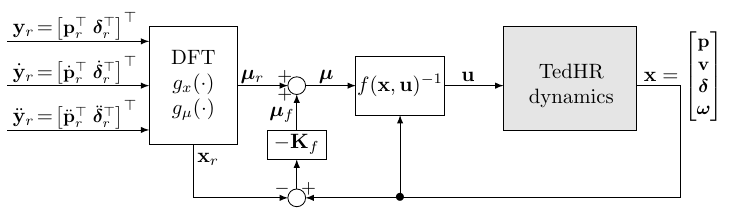}
   \caption{Flatness-based Controller  (FC) architecture}
   \label{fig:FCstructure}
\end{figure}

The FC structure is depicted in Figure~\ref{fig:FCstructure}. 
The controller inputs consist of the references $\mathbf{y}_r=\colvec{\mathbf{p}_r^\top \; \boldsymbol{\delta}_r^\top}^\top \in \mathbb{R}^6$, and their first and second derivatives. The \textit{DFT - Differential Flatness Transformation} block computes the reference state $\mathbf{x}_r \in \mathbb{R}^{12}$ and reference flat input $\boldsymbol{\mu}_r \in\mathbb{R}^6$ by resorting on the functions $g_x(\cdot)$ and $g_\mu(\cdot)$ previously introduced. Then, $\mathbf{x}_r$ is used to compute the feedback action $\boldsymbol{\mu}_{f} \in \mathbb{R}^6$. Formally, it is  $\boldsymbol{\mu}_f = - \mathbf{K}_f (\mathbf{x}-\mathbf{x}_r)$ where the gain matrix $\mathbf{K}_f \in \mathbb{R}^{6 \times 12}$ is computed adopting the LQR approach on system~\eqref{eqn:model:ss:full}. The resulting vector $\boldsymbol{\mu}=\boldsymbol{\mu}_r + \boldsymbol{\mu}_f$ is then converted in terms of (dynamics) input vector $\mathbf{u}$ by inverting the relation~\eqref{eqn:model:ss:input}, which requires the state feedback.

%%%%%%%%%%%%%%%%%%%%%%%%%%%%%%%%%%%%%%%%%%%%%%%%%%%%%%%%%%%%%%%%%%%%%%%%%%%%%%%%

\begin{figure*}[t!]
    \centering
\includegraphics[width=0.98\textwidth]{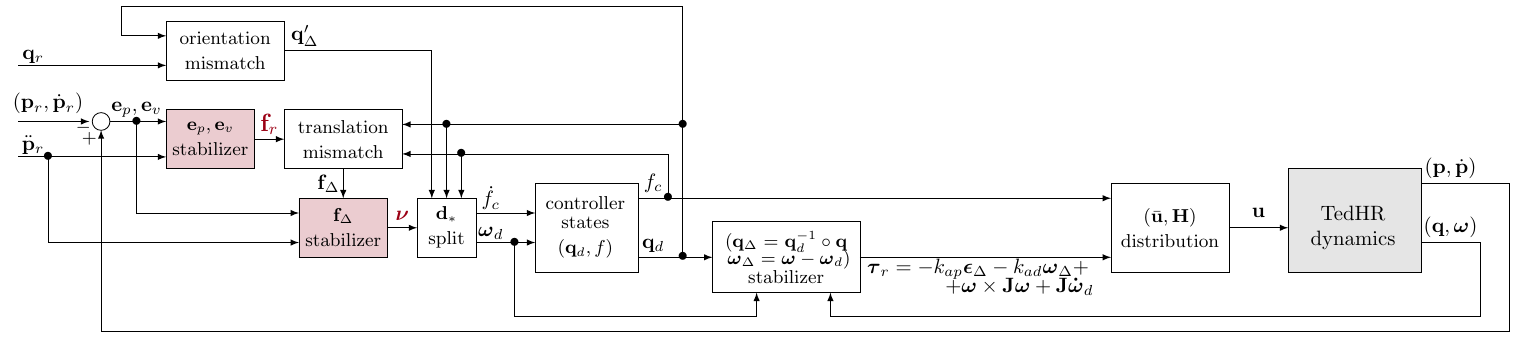}
    \caption{Hierarchical Controller (HC) architecture}
    \label{fig:HCstructure}
\end{figure*}

\section{HIERARCHICAL CONTROLLER}
\label{sec:hierarchical}

The proposed HC constitutes an extension of the nonlinear control approach described in~\cite{michieletto2017nonlinear,michieletto2020hierarchical}. This is based on the fulfillment for the TedHR control input matrices of the algebraic condition $\emph{rk}(\mathbf{M}\bar{\mathbf{F}}) = 3$, where $\bar{\mathbf{F}} \in 
\mathbb{R}^{6 \times 3}$ is so that $\mathrm{Im}(\bar{\mathbf{F}})=\ker(\mathbf{F})$. Such a condition guarantees the existence of a matrix $\mathbf{H} \in \mathbb{R}^{6 \times 6}$ such that $\mathbf{M}\mathbf{H}\mathbf{M}^\top$ is invertible and $\mathbf{F}\mathbf{M}^\dagger_\mathbf{H} = \mathbf{0}_3$, where $\mathbf{M}^\dagger_\mathbf{H}= \mathbf{H}\mathbf{M}^\top (\mathbf{M}\mathbf{H}\mathbf{M}^\top)^{-1} \in \mathbb{R}^{6 \times 3}$ is the generalized right pseudo-inverse of $\mathbf{M}$. 
The HC input is then designed as
\begin{equation}
\label{eq:HC:input}
\mathbf{u}=\mathbf{M}_{\mathbf{H}}^{\dagger} \boldsymbol{\tau}_{r}+\bar{\mathbf{u}} f_c
\end{equation}
where $\boldsymbol{\tau}_{r} \in \mathbb{R}^{3}$ is the reference moment and $f_c \in \mathbb{R}$ is the control force intensity, namely $f_c= \Vert \mathbf{f}_c \Vert$. The vector $\bar{\mathbf{u}} \in \mathbb{R}^{6}$ is selected in $\ker(\mathbf{M})$ so that the product $\mathbf{F} \bar{\mathbf{u}}$  identifies a direction in the force space $\text{Im}(\mathbf{F}) \cap \mathbb{S}^2$, referred to as \textit{zero-moment preferential direction} $\mathbf{d}_\ast$, along which the control force can be independently assigned with respect to the control moment. Note that, based on~\eqref{eq:control_force_moment}, the input~\eqref{eq:HC:input} implies $\mathbf{f}_{c}=\mathbf{F} \mathbf{u}=\mathbf{d}_{*} f_c$ and $\boldsymbol{\tau}_{c}=\mathbf{M u}=\boldsymbol{\tau}_{r}$.

To address Problem~\ref{prb:almost6Dcontrol}, we observe that a suitable selection of $\bar{\mathbf{u}}$ is such that the resulting zero-moment preferential direction corresponds to
\begin{equation}
\label{eq:pref_direction}
    \mathbf{d}_{*} = {\mathbf{d}}/{\Vert \mathbf{d}\Vert }, \quad \mathbf{d}=\mathbf{R}\left(\mathbf{q}_{r}\right)^\top \left(mg\mathbf{e}_3 + m\mathbf{\ddot{p}}_r\right). 
\end{equation}
The choice~\eqref{eq:pref_direction}, indeed, entails that the resulting control force $\mathbf{f}_c = \mathbf{d}_{*} f_c $ is oriented, in the body frame, in order to counterbalance the gravity force while acting along the direction of the reference position trajectory.

The HC structure is reported in Figure~\ref{fig:HCstructure}. The controller states are the control force intensity $f_c$ and the \textit{desired orientation} $\mathbf{q}_{d} \in \mathbb{S}^{3}$. This latter is the rotation that ensures the zeroing of the \textit{force mismatch} vector $\mathbf{f}_\Delta \in \mathbb{R}^3$ defined as the difference between the \textit{desired control force}  $\mathbf{R}\left(\mathbf{q}_{d}\right) \mathbf{d}_{*}f_c$ and the \textit{reference force} $\mathbf{f}_r \in \mathbb{R}^3$, both expressed in the world frame. We remark that the definition of $\mathbf{f}_r$ is revised with respect to~\cite{michieletto2020hierarchical} since we deal with the trajectory tracking problem, rather than with the static hovering stabilization. Aiming at steering the platform along the reference position profile while counterbalancing gravity, we choose
\begin{equation}
\label{eq:HC:f_r}
     \mathbf{f}_r = mg\mathbf{e}_3 +m \mathbf{\ddot{p}}_r - k_{pp} \mathbf{e}_p - k_{pd} \mathbf{e}_v, 
\end{equation}
where $\mathbf{e}_{p}=\mathbf{p}-\mathbf{p}_{r} \in \mathbb{R}^{3}$ and $\mathbf{e}_{\dot{p}}=\mathbf{\dot{p}} - \mathbf{\dot{p}}_{r} \in \mathbb{R}^{3}$ are the position and velocity error vectors, and  $k_{p p}, k_{p d} \in \mathbb{R}$ are the corresponding tunable positive scalar gains. 

As proven in~\cite{michieletto2020hierarchical}, zeroing the force mismatch vector $\mathbf{f}_\Delta$ implies the stabilization of the TedHR translational dynamics. In particular, this is possible by imposing 
\begin{subequations}
\label{eq:HC:omega_d_dot_f}
   \begin{align}
        \boldsymbol{\omega}_{d}&= \boldsymbol{\omega}_{d}^{0} + \boldsymbol{\omega}_{d}^{\prime} = \frac{1}{f}\left[\mathbf{d}_{*}\right]_{\times} \mathbf{R}\left(\mathbf{q}_{d}\right)^{\top} \boldsymbol{\nu} -k_q \mathbf{d}_* 
 \mathbf{d}_*^\top \boldsymbol{\epsilon}_\Delta^\prime \\
        \dot{f}_c&=\left(\mathbf{R}\left(\mathbf{q}_{d}\right) \mathbf{d}_{*}\right)^{\top} \boldsymbol{\nu}  
   \end{align} 
\end{subequations}
where the formulation of the $\boldsymbol{\omega}_{d} \in \mathbb{R}^3$ highlights a twofold action, regulated by the positive scalar gain $k_q \in \mathbb{R}$. 
On the one hand, by means of $\boldsymbol{\omega}_{d}^{0} \in \mathbb{R}^3$, it accommodates the platform orientation along the desired position trajectory; on the other, through $\boldsymbol{\omega}_{d}^{\prime} \in \mathbb{R}^3$, it regulates the dynamics of the controller state $\mathbf{q}_d$ towards $\mathbf{q}_r$ by acting on an orientation mismatch term $\mathbf{q}_\Delta^\prime\in \mathbb{S}^3$, $\mathbf{q}_\Delta^\prime= \colvec{ \eta_\Delta^\prime \; \boldsymbol{\epsilon}_\Delta^\prime }^\top=\mathbf{q}_r^{-1} \circ \mathbf{q}_d  $ between the reference and the desired orientations. 
The vector $\boldsymbol{\nu} \in \mathbb{R}^3$ appearing in~\eqref{eq:HC:omega_d_dot_f} constitutes an additional virtual input of the controller whose selection is modified as compared to~\cite{michieletto2020hierarchical}. Indeed, consequently to~\eqref{eq:HC:f_r}, we set 
 \begin{equation}
 \label{eq:HC:nu}
\boldsymbol{\nu} = \frac{k_{p d} k_{p p}}{m} \mathbf{e}_{p} + \left(\frac{k_{p d}^{2}}{m}-k_{p p}\right) \mathbf{e}_{v}  -\left(\frac{k_{p d}}{m} + k_\Delta \right) \mathbf{f}_{\Delta} -m \mathbf{\dddot{p}}_r,
\end{equation}
with $k_{\Delta} \in \mathbb{R}$ being an additional positive scalar gain. 

\begin{remark}
    Assuming that the UAV orientation $\mathbf{q}$ has converged to the reference $\mathbf{q}_r$, the choices~\eqref{eq:HC:omega_d_dot_f} and~\eqref{eq:HC:nu} ensure that $\dot{\mathbf{f}}_{\Delta}=-k_{\Delta} \mathbf{f}_{\Delta}$. Then, also the force mismatch converges to zero, and the position tracking is fulfilled. Indeed, recalling that $\mathbf{f}_{\Delta} = \mathbf{R}\left(\mathbf{q}_{d}\right) \mathbf{d}_{*}f_c - \mathbf{f}_r$ with $\mathbf{f}_r$ as in~\eqref{eq:HC:f_r}, it is possible to verify that
\begin{subequations}
 \begin{align}
    \dot{\mathbf{f}}_{\Delta} & =\mathbf{R}\left(\mathbf{q}_{d}\right) \mathbf{d}_{*} \dot{f}_c+\dot{\mathbf{R}}\left(\mathbf{q}_{d}\right) \mathbf{d}_{*} f-\dot{\mathbf{f}}_{r} \\
&=\dot{\mathbf{f}}_{\Delta, 1}+\dot{\mathbf{f}}_{\Delta, 2}+\dot{\mathbf{f}}_{\Delta, 3}
\end{align}       
\end{subequations}
with 
\begin{subequations}
\begin{align}
 \dot{\mathbf{f}}_{\Delta, 1}  &=\mathbf{R}\left(\mathbf{q}_{d}\right) \mathbf{d}_{*} \mathbf{d}_{*}^{\top} \mathbf{R}\left(\mathbf{q}_{d}\right)^{\top} \boldsymbol{\nu}   \\
  \dot{\mathbf{f}}_{\Delta, 2} & = \boldsymbol{\nu} - \mathbf{R}\left(\mathbf{q}_{d}\right) \mathbf{d}_{*} \mathbf{d}_{*}^{\top} 
 \mathbf{R}\left(\mathbf{q}_{d}\right)^{\top} \boldsymbol{\nu} \\
 \dot{\mathbf{f}}_{\Delta, 3} & = k_{p p} \mathbf{e}_{v} \!+\!\frac{k_{p d}}{m}\left( -k_{p p} \mathbf{e}_{p}-k_{p d} \mathbf{e}_{v}+\mathbf{f}_{\Delta}\right) +m \mathbf{\dddot{p}}_r
 \end{align}
\end{subequations}
%{\color{red} ...}
\end{remark}

The HC architecture in Figure~\ref{fig:HCstructure} is completed by an appropriate selection of the reference control moment $\boldsymbol{\tau}_r \in \mathbb{R}^3$ that ensures the zeroing of $\mathbf{q}_\Delta= \colvec{ \eta_\Delta\; \boldsymbol{\epsilon}_\Delta }^\top=\mathbf{q}_d^{-1} \circ \mathbf{q} \in \mathbb{S}^3$. 
Since the control of the rotational dynamics and the realization of $\boldsymbol{\tau}_r$ are as in~\cite{michieletto2020hierarchical}, we do not report here the detailed derivation. Nonetheless, we point out the following fact.

\begin{remark}
The designed controller guarantees the regulation of the UAV orientation towards the desired one (internal controller state), rather than the reference one (external controller input), thus highlighting the lower priority assigned to attitude regulation in the hierarchical architecture.
\label{rem:positionpriority}
\end{remark}

%%%%%%%%%%%%%%%%%%%%%%%%%%%%%%%%%%%%%%%%%%%%%%%%%%%%%%%%%%%%%%%%%%%%%%%%%%%%%%%%

\section{VALIDATION}
\label{sec:validation}

{To assess the performance of both the FC and HC solutions, we account for a TedHR required to track a circular reference path while concurrently adjusting its orientation in 3D space. 
Specifically, $\mathbf{p}_r$ is designed as a circular trajectory with radius $\SI{2}{m}$ at a constant altitude of $\SI{1}{m}$ from the ground and a set of steps is imposed on the components of $\boldsymbol{\delta}_r$. 
In detail, motivated by the UAV planar structure on the $xy$ plane, the reference roll and pitch angles $\phi_r$ and $\theta_r$ are defined as a step sequence of respectively $\colvec{ -7^\circ \; 0^\circ \; 7^\circ}$ and $\colvec{ 0^\circ \; 3.5^\circ \; 7^\circ}$, while the reference yaw angle $\psi_r$ is designed to be more aggressive, varying in the range $\colvec{90^\circ, 270^\circ}$ with increasing steps of $45^\circ$. 
The initial conditions for such a task are those of ground parking (i.e., with zero pose and velocities), therefore the first phase in the proposed scenario involves a take-off action.}

\def\figureheight{2.8cm}
 \begin{figure*}[t!]
  \centering
          \includegraphics[width=1\textwidth]{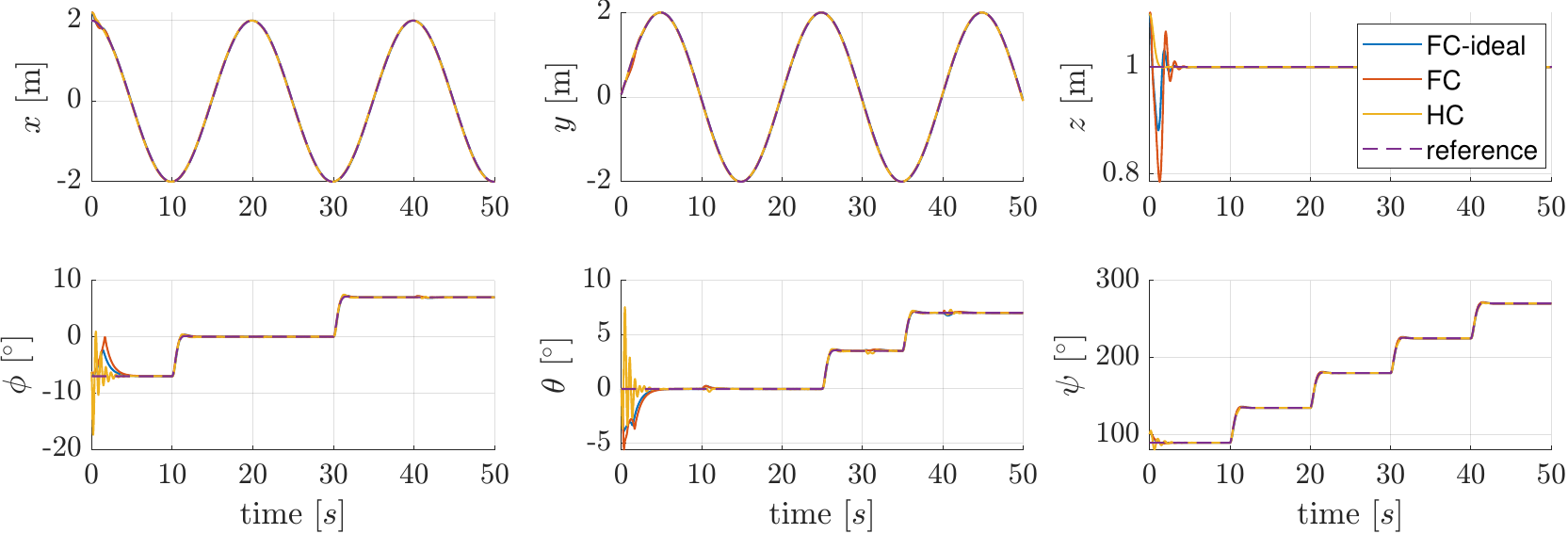}
%          \includegraphics[height=\figureheight]{images/sims_lots_of_steps/img_trj_x_no_noise.pdf}
% \hfill
%          \includegraphics[height=\figureheight]{images/sims_lots_of_steps/img_trj_y_no_noise.pdf}
% \hfill
%          \includegraphics[height=\figureheight]{images/sims_lots_of_steps/img_trj_z_no_noise.pdf}
% \\
%          \includegraphics[height=\figureheight]{images/sims_lots_of_steps/img_trj_phi_no_noise.pdf}
% \hfill
%          \includegraphics[height=\figureheight]{images/sims_lots_of_steps/img_trj_theta_no_noise.pdf}
% \hfill
%          \includegraphics[height=\figureheight]{images/sims_lots_of_steps/img_trj_psi_no_noise.pdf}
%  \caption{position and attitude regulation error - \textbf{unperturbed} flight conditions  (scenario A)}
  \caption{{Scenario A: \textbf{unwindy} flight conditions. Position (top row) and attitude (bottom row) behaviors for the three control architectures: FC-ideal, FC, HC.}}
  \label{fig:unperturbed}
 \end{figure*}%

The simulation is performed in the MATLAB-Simulink environment by modeling several real-world nonidealities. 
We consider a star-shaped tilted hexarotor with diameter of $\SI{\sim0.8}{m}$ (propellers included), 
 mass of $\SI{\sim3.5}{kg}$ and tilt angles set to $\alpha=\SI{25}{deg}$ and $ \beta=\SI{10}{deg}$ (guaranteeing the existence of the matrix $\mathbf{H}$ involved in the definition of the HC input~\eqref{eq:HC:input}). 
For both the FC and the HC, the feedback signals of position and orientation and their derivatives are affected by a time delay $t_f =\SI{12}{m}s$ and additive Gaussian noise with zero mean and variance as illustrated in Table~\ref{tab:sim_params}. Moreover, 
the UAV state is made available to the controller at %lower sampling frequency of 
$\SI{100}{Hz}$, according to  the features of a typical %motion capture system (for indoor use) and moreover to those of a 
IMU sensor; the propellers spinning rates,  taking action in the definition of the entries of the vector $\mathbf{u}$, are bounded in $[0 \; 83.5]\SI{}{Hz}$.

%\begin{remark}
%The considered UAV is supposed to be equipped with a camera whose focal axis is aligned with $\mathbf{y}_B$. Thus, in the desired pointing task, it has to track the reference 
%\vspace{-0.1cm}
%in~\eqref{eqn:trj:p}, where the orientation $\bm \delta_r$ is calculated such that the camera points to the POI from the current position of the UAV.
%{\color{red} \begin{subequations}
%\begin{align}
%\label{eqn:trj:p}
%    \mathbf{p}_r &= \colvec{p_{x,r} \; p_{y,r} \; p_{z,r}}^\top 
%    =
%    \colvec{
%    2 \sin(\frac{2 \pi t}{10}) & 2 \cos(\frac{2 \pi t}{10}) & 1}^\top \\
%    \boldsymbol{\delta}_r &= \colvec{\phi_r \; \theta_r \; \psi_r}^\top = \colvec{-5.71 & 0 & (36t+90)}^\top
%\end{align}
% \end{subequations}}
% where the resulting expression of $\boldsymbol{\delta}_r$ depends on the (fixed) camera orientation in $\mathscr{F}_B$ and on the given POI.  
%
%  When coping with a pointing task, the employed UAV is assumed to be equipped with a camera having a fixed position and orientation in $\mathscr{F}_B$. Introducing the reference camera frame $\mathscr{F}_C=\{O_C, (\mathbf{x}_C, \mathbf{y}_C, \mathbf{z}_C)\}$, the 6D trajectory to track is usually defined so that the focal axis of the camera, coinciding with $\mathbf{z}_C$, is constantly directed towards a point of interest (PoI) $P_d$ which is also the center the circular maneuver required to the aerial platform. 
%\end{remark}

\begin{table}[b]
    \centering
    \begin{tabular}{l||ccc}
    \toprule
    & $x$ component & $y$ component & $z$  component\\
            \midrule
        $\mathbf{p}$ [m$^2$]  &
        $4.099*10^{-7}$&$2.838*10^{-7}$&$2.105*10^{-8}$ \\
        $\boldsymbol{\delta}$ [deg$^2$] & 
        $0.0012$ & $0.0011$ & $0.0011$ \\
        $\mathbf{v}$ [(m/s)$^2$] & 
        $2.050*10^{-6}$  &  $1.419*10^{-6}$&   $1.050*10^{-7}$ \\
        $\boldsymbol{\omega}$ [(deg/s)$^2$] &
        $0.0024$ & $0.0022$ & $0.0022$ \\
        \bottomrule
    \end{tabular}
    \caption{simulation parameters -  noise variance}
    \label{tab:sim_params}
\end{table}

To comparatively evaluate the control architectures discussed in Sections~\ref{sec:flatness}-\ref{sec:hierarchical}, two different scenarios are taken into account:
\begin{enumerate}
    \item[A.] \textit{unwindy flight conditions} - the considered TedHR platform is required to fulfill the described task under the given assumptions about the signal delay and observation noise models (but without wind disturbances);
    \item[B.] \textit{windy flight conditions} - we perform the tests by adding further disturbances induced by the wind action.
\end{enumerate}
As regards the wind action, we consider a shear component and a Dryden turbulence component, both of them directed as $\mathbf{x}_W$ and such that the wind speed is equal to {$\SI{10}{m/s}$} at $\SI{6}{m}$ of altitude. In addition, we also model the occurrence of a wind gust at $T_w=\SI{25}{s}$ in order for its velocity {(along the three world frame directions)} to be equal to {$\colvec{2\; 2\; 1}^\top \SI{}{m/s}$} after $\SI{2}{s}$. 
The effect of the whole wind action results in an adverse force $\mathbf{f}_w \in \mathbb{R}^3$ proportional to UAV invested area $A \in \mathbb{R}$. Formally, it is $\mathbf{f}_w =  \rho A \mathbf{d}_w$, 
where $\rho \in \mathbb{R}$ is the (time-varying) air pressure coefficient defined according to the COESA atmosphere model and dependent on the vehicle altitude, and $\mathbf{d}_w \in \mathbb{R}^3$ is the difference between the wind velocity resulting from all its components and the UAV velocity. Then, $A$ is estimated as $A = (1-d_v)A_\ell + d_v A_u$, where $A_\ell=\SI{0.111}{m^2}$ and $A_u=\SI{0.885}{m^2}$ respectively approximate the lateral and upper area of the considered TedHR platform and $d_v \in \mathbb{R}$ is computed as $d_v = ({\mathbf{d}_w}/{\|\mathbf{d}_w\|}) \cdot \mathbf{z}_B$.

The performance of the FC and HC in scenarios A and B are also compared to an ideal situation, where the FC solution is adopted, and no delay or observation noise is affecting the dynamics. In this case (named \emph{FC-ideal}), the feedback control component has to compensate only for the initial conditions mismatch: this represents almost the best possible solution to the position tracking and attitude regulation problem given the system dynamics and constraints.

For validation, we perform 500 Monte-Carlo (MC) simulations in both A and B scenarios. In the remainder of this section, we discuss the achieved results by showing in figures some representative MC evolutions and by evaluating the following performance indexes in a summary table:
\begin{itemize}[leftmargin=0.5cm]
    %\item the \textit{position tracking error}  $\mathbf{e}_p = \mathbf{p}_r-\mathbf{p} \in \mathbb{R}^3$;
    \item {the \textit{position tracking error norm}  $e_p = \|\mathbf{e}_p\| \in \mathbb{R}$};
    \item the \textit{attitude tracking error} ${e}_a
= 2\arccos(\mathbf{q}^T \mathbf{q}_r) \in \mathbb{R}$, computed as the Riemannian geodesic distance on $\mathbb{S}^3$; 
\item the \textit{control input norm} $u_n = \Vert \mathbf{u} \Vert \in \mathbb{R}$, which provides an intuition on the controller energy consumption;
\item the \textit{control input excess} $u_e \in \mathbb{R}$ defined in case of input saturation as the difference between the maximum value among the unbounded entries of $\mathbf{u}$ and the upper limit of $\SI{83.5}{Hz}$. This gives an insight into the control feasibility. 
\end{itemize}

 % FIGs SCENARIO A
Figure~\ref{fig:unperturbed} reports the pose trend in a single representative test for scenario A. 
In the case of unwindy flight conditions, the FC and the HC perform pretty similarly in position tracking and attitude regulation: both the reference position and the reference orientation are followed by ensuring very small (if not zero) steady-state errors.
To get further insight into these results, we refer also to Figure~\ref{fig:unperturbed_delta} in the Appendix, where the mismatch between the UAV position and orientation and the corresponding references is reported.

\def\figureheight{2.8cm}
\begin{figure*}[t!]
  \centering
          \includegraphics[width=\textwidth]{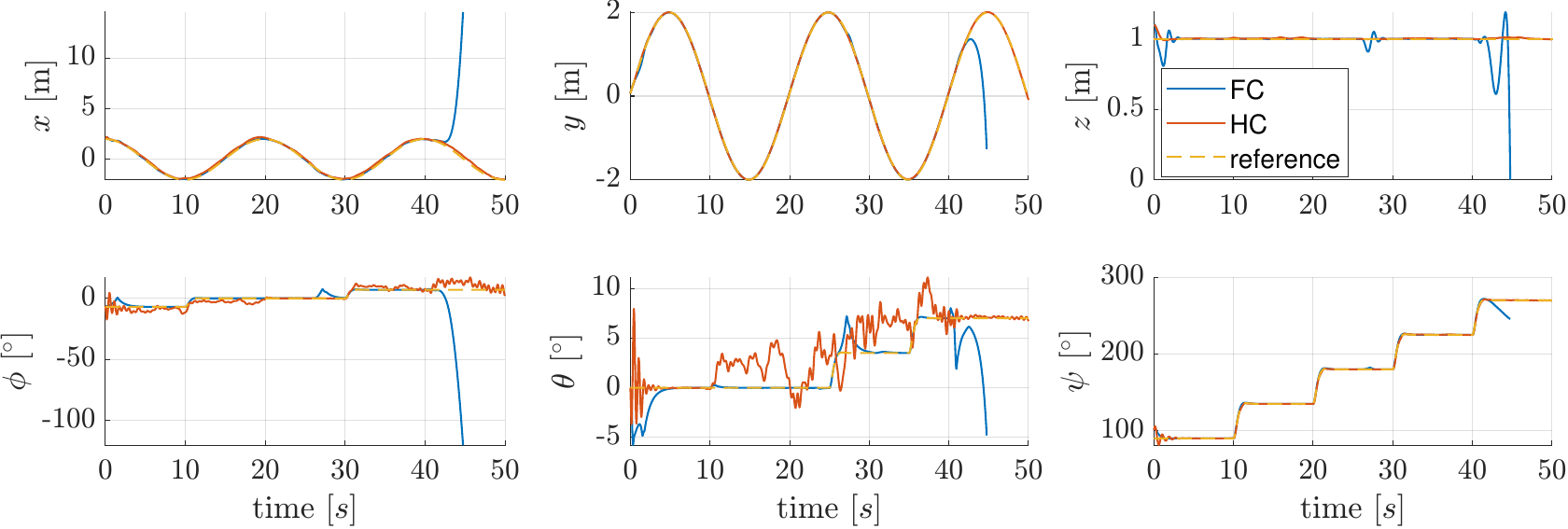}
%          \includegraphics[height=\figureheight]{images/sims_lots_of_steps/img_trj_x_noise.pdf}
% \hfill
%          \includegraphics[height=\figureheight]{images/sims_lots_of_steps/img_trj_y_noise.pdf}
% \hfill
%          \includegraphics[height=\figureheight]{images/sims_lots_of_steps/img_trj_z_noise.pdf}
% \\
%          \includegraphics[height=\figureheight]{images/sims_lots_of_steps/img_trj_phi_noise.pdf}
% \hfill
%          \includegraphics[height=\figureheight]{images/sims_lots_of_steps/img_trj_theta_noise.pdf}
% \hfill
%          \includegraphics[height=\figureheight]{images/sims_lots_of_steps/img_trj_psi_noise.pdf}
 \caption{{Scenario B: \textbf{windy }flight conditions. Position (top row) and attitude (bottom row) behaviors for the two control architectures: FC, HC.}}
 \label{fig:perturbed}
\end{figure*}%

\begin{figure*}[b!]
          \includegraphics[width=0.98\textwidth]{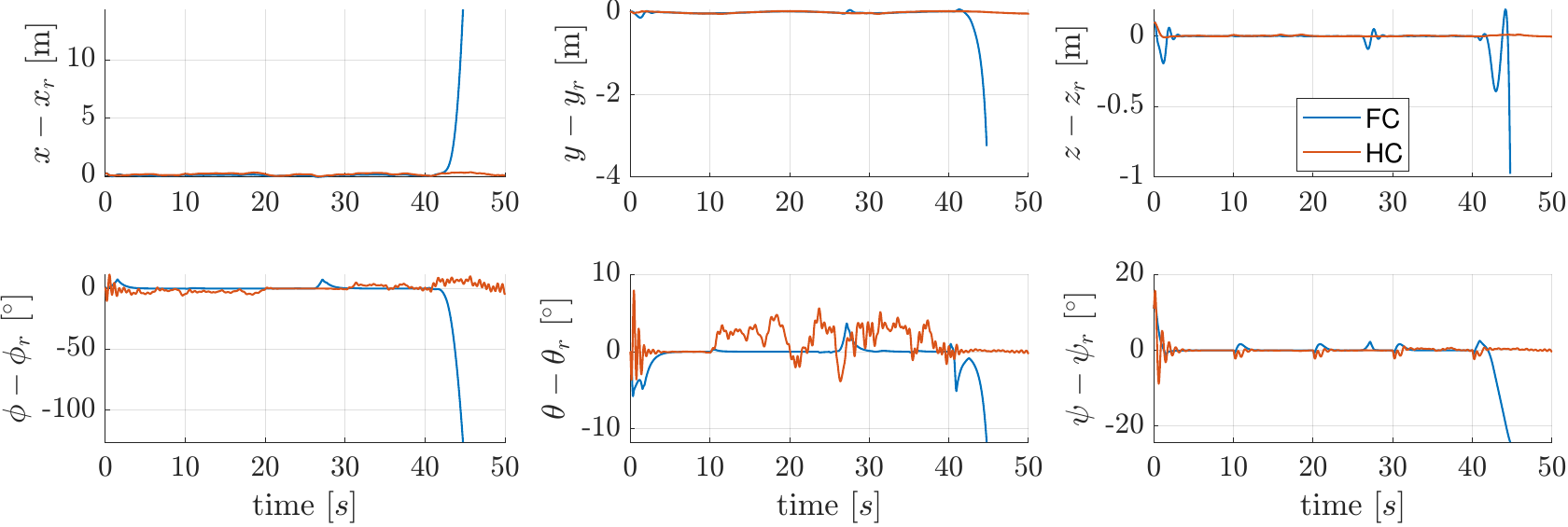}
%   \centering
%          \includegraphics[height=\figureheight]{images/sims_lots_of_steps/img_err_x_noise.pdf}
% \hfill
%          \includegraphics[height=\figureheight]{images/sims_lots_of_steps/img_err_y_noise.pdf}
% \hfill
%          \includegraphics[height=\figureheight]{images/sims_lots_of_steps/img_err_z_noise.pdf}
% \\
%          \includegraphics[height=\figureheight]{images/sims_lots_of_steps/img_err_phi_noise.pdf}
% \hfill
%          \includegraphics[height=\figureheight]{images/sims_lots_of_steps/img_err_theta_noise.pdf}
% \hfill
%          \includegraphics[height=\figureheight]{images/sims_lots_of_steps/img_err_psi_noise.pdf}
%  \caption{position and attitude regulation error - \textbf{unperturbed} flight conditions  (scenario A)}
  \caption{Scenario B: \textbf{windy }flight conditions. Position and attitude mismatch along the components $\colvec{x-x_r \; y-y_r \; z-z_r}$ (top row) and $\colvec{\phi-\phi_r \; \theta-\theta_r \; \psi-\psi_r}$ (bottom row) for the three control architectures: FC, HC.}
  \label{fig:perturbed_delta}
 \end{figure*}

 % FIGs SCENARIO B
{The situation is completely different when the flight is affected by wind disturbances. 
Figure~\ref{fig:perturbed} reports the position and orientation trends for scenario B, and, also in this case, a detailed view of the pose mismatch is given in Figure~\ref{fig:perturbed_delta}. 
From these results, it can be appreciated how, with respect to the unwindy case, the performance of the HC is almost invariant as regards the accuracy of the position tracking and remains reasonably close to the reference,  converging to steady-state zero error, for the attitude regulation.
Conversely, the FC solution turns out to be not robust in the presence of the wind gust action: both the position and attitude regulation error diverge after just over $\SI{42}{s}$.}

\begin{table}[b]
    \centering
    \begin{tabular}{l||c||cc|cc}
    \toprule
    & FC-ideal & FC-A & HC-A & FC-B & HC-B \\
    \midrule
$e_p$ [m]   & $0.010$ & $0.042$ & $0.041$ & -- & $0.064$ \\ % FC-B $0.042$
$e_a$ [deg] & $0.466$ & $0.556$ & $0.439$ & -- & $1.352$ \\ % FC-B $0.618$
$u_n$ [Hz]  & $4011$ & $4007$ & $4012$ & -- & $4012$ \\ % FC-B $4021$
$u_e$ [Hz]  & $0.722$ & $2.052$ & $0.118$ & -- & $0.120$ \\ % FC-B $1.822$
%
%
% $e_p$ [m]     & $   0.065 $         & $   0.215 $           & $ 0.318 $           & $  0.212      $ \\
% $e_a$ [deg]   & $   4.127 $         & $   36.59 $           & $ 4.691 $           & $  36.75      $ \\
% $u_n$ [Hz]    & $ \phantom{.}4102 $ & $   \phantom{.}4103 $ & $ \phantom{.}4035 $ & $  \phantom{.}4102     $ \\
% $u_e$ [Hz]    & $0.054 $            & $0.141 $              & $ 2.759 $           & $  0.146$ \\
%    noise thr [N]           & $ 1.367  $      & $        -    $ & $ 3.733      $ & $ -            $ \\
%    $u_n$ -total                  & $ 2.051\cdot10^5 $  & $  2.051\cdot10^5 $ & $ 1.879\cdot10^5 $ & $  2.051\cdot10^5  $ \\

 %   $e_a$-tot [deg]    & $   \phantom{0}60.8  $     & $       5067  $ & $ \phantom{0}58.4       $ & $    5114      $ \\

 %   $e_p$-final [m]      & $   0.055 $     & $       0.180 $ & $ 3.175      $ & $   0.187      $ \\
    \bottomrule
    \end{tabular}
    \caption{controllers performance indexes}
    \label{tab:kpis}
\end{table}

% SUMMARY TABLE
{To complement these findings, \autoref{tab:kpis} reports the mean values of the performance indexes computed on all the MC trails.} 
{Focusing on the second and third columns, we note that FC and HC exhibit similar average performance as for the pose control in scenario A.
We also remark that the two control approaches are equivalent in terms of energy consumption since the index $u_n$ is similar in correspondence to the two controllers, nonetheless, the mean value of the control input excess is much smaller for the HC suggesting a higher level of feasibility.
When it comes to the analysis of scenario B, we report only the results for the stable HC architecture, confirming the accuracy in position tracking at the cost of an overall worsening of the attitude regulation performance in the presence of wind action.
This observation is in line with what is stated in Remark~\ref{rem:positionpriority}.
As a final comment, we can observe that the lack of robustness of the FC architecture can be suggested by the high mean value of the $u_e$ index: in correspondence to FC, this is an order of magnitude higher as compared to the HC case for which, instead, it remains almost the same of the unwindy scenario. 
}

 \begin{figure}[b!]
  \centering
         \includegraphics[width=0.75\columnwidth, trim = {0.0cm 0.65cm 0.0cm 0},clip]{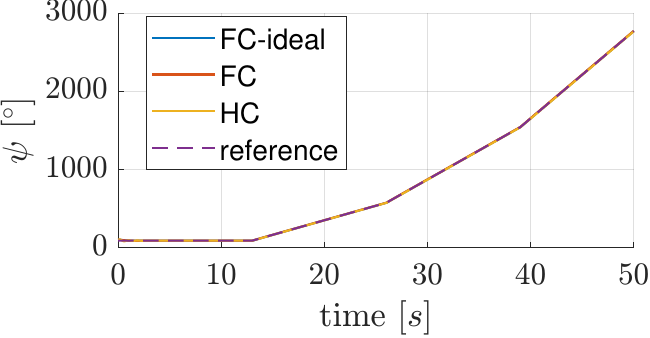}{\phantom{A}}\\ \vspace{0.25cm}
         \includegraphics[width=0.72\columnwidth]{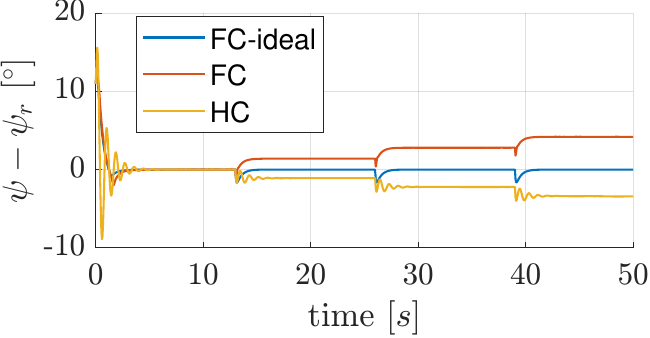}
%          \includegraphics[height=\figureheight]{images/sim_ramps/img_err_x_no_noise.pdf}
% \hfill
%          \includegraphics[height=\figureheight]{images/sim_ramps/img_err_y_no_noise.pdf}
% \hfill
%          \includegraphics[height=\figureheight]{images/sim_ramps/img_err_z_no_noise.pdf}
% \\
%          \includegraphics[height=\figureheight]{images/sim_ramps/img_err_phi_no_noise.pdf}
% \hfill
%          \includegraphics[height=\figureheight]{images/sim_ramps/img_err_theta_no_noise.pdf}
% \hfill
%          \includegraphics[height=\figureheight]{images/sim_ramps/img_err_psi_no_noise.pdf}
  \caption{Scenario C: {\bf unwindy} flight conditions with angle ramp reference. $\psi$ (top) and $\psi-\psi_r$ (bottom) behaviors for the three control
architectures: FC-ideal, FC, HC.}
  \label{fig:unperturbed_ramp}
 \end{figure}%

\def\figureheight{2.8cm}
 \begin{figure*}[b!]
  \centering
          \includegraphics[width=0.98\textwidth]{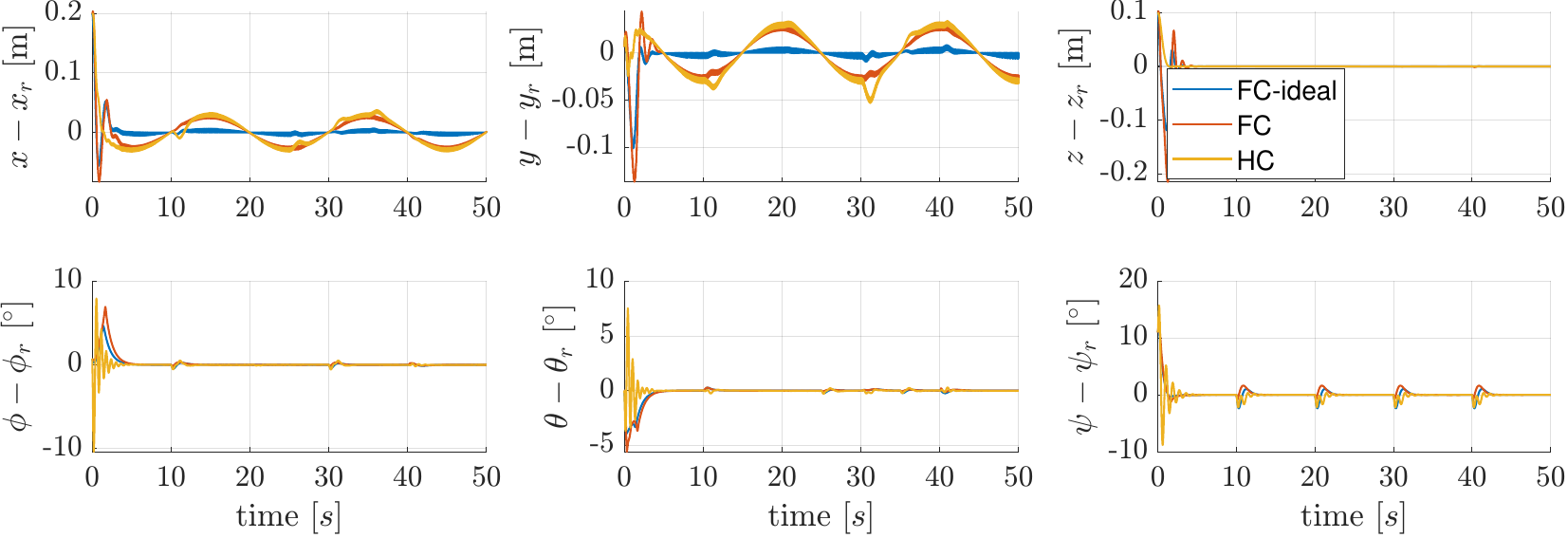}
%          \includegraphics[height=\figureheight]{images/sims_lots_of_steps/img_err_x_no_noise.pdf}
% \hfill
%          \includegraphics[height=\figureheight]{images/sims_lots_of_steps/img_err_y_no_noise.pdf}
% \hfill
%          \includegraphics[height=\figureheight]{images/sims_lots_of_steps/img_err_z_no_noise.pdf}
% \\
%          \includegraphics[height=\figureheight]{images/sims_lots_of_steps/img_err_phi_no_noise.pdf}
% \hfill
%          \includegraphics[height=\figureheight]{images/sims_lots_of_steps/img_err_theta_no_noise.pdf}
% \hfill
%          \includegraphics[height=\figureheight]{images/sims_lots_of_steps/img_err_psi_no_noise.pdf}
%  \caption{position and attitude regulation error - \textbf{unperturbed} flight conditions  (scenario A)}
  \caption{Scenario A: \textbf{unwindy} flight conditions. Position and attitude mismatch along the components $\colvec{x-x_r \; y-y_r \; z-z_r}$ (top row) and $\colvec{\phi-\phi_r \; \theta-\theta_r \; \psi-\psi_r}$ (bottom row) for the three control architectures: FC-ideal, FC, HC.}
  \label{fig:unperturbed_delta}
 \end{figure*}%

\def\figureheight{2.8cm}

We conclude this analysis by presenting a further unwindy scenario the attitude is dynamically changing with a ramp reference (scenario C). In detail, we impose constant reference roll and pitch angles while the yaw reference angle is designed to change with ramps of different slopes. Interestingly, this situation translates into a steady-state error in the attitude regulation task, which turns out to be proportional to the ramp slope. This fact is shown in Figure~\ref{fig:unperturbed_ramp}.
{We conjecture that this behavior is related to the imposed constant yaw rate, which is not taken into account since the feedback control scheme does not consider an angular rate reference, differently from what is done with the translational dynamics with the velocity reference. 
Current activity is ongoing to get further insights into this behavior and to devise a suitable control action to compensate for the attitude errors.}

%%%%%%%%%%%%%%%%%%%%%%%%%%%%%%%%%%%%%%%%%%%%%%%%%%%%%%%%%%%%%%%%%%%%%%%%%%%%%%%%

\section{CONCLUSIONS}
\label{sec:conclusions}

In this paper, we present two controllers for a TedHR platform required to track a position trajectory while attaining a concurrent attitude regulation with respect to step references. The first one (FC) relies on the differential flatness property of the considered platform to design a suitable feedforward control action ensuring the tracking of both a position and attitude reference profile in conjunction with a LQR scheme acting in feedback. 
The other one (HC) is a nonlinear hierarchical regulator having a cascaded structure wherein the orientation reference is tracked with lower priority. 

The performance of the two controllers is compared in both unwindy and windy scenarios in a MATLAB-Simulink environment. We verify that the FC architecture, which represents also a reference benchmark, stands out for the tracking accuracy in the unwindy case, thus encouraging its exploitation in indoor applications. On the other side, the proposed HC solution extends the static hovering controller described in~\cite{michieletto2020hierarchical} allowing to reach good tracking and regulation performances even in the presence of external disturbance. Indeed, contrarily to the FC, the HC turns out to be robust when the wind action is taken into account, simulating the typical (more challenging) outdoor conditions.

{Future work includes the improvement of the HC architecture in order to control full 6D trajectories, where the entire pose (position and attitude) is concurrently tracked, also devising some strategy to mitigate the effect of the positioning priority over the orientation regulation.}

%%%%%%%%%%%%%%%%%%%%%%%%%%%%%%%%%%%%%%%%%%%%%%%%%%%%%%%%%%%%%%%%%%%%%%%%%%%%%%%%
\appendix
%\subsection{additional plots pose error}

{In this Appendix, we report additional plots highlighting the behavior of the control architectures in Scenario A and B. 
In Figures~\ref{fig:unperturbed_delta}-\ref{fig:perturbed_delta}, an insight into the position and attitude mismatch is shown.}

\vfill

%%%%%%%%%%%%%%%%%%%%%%%%%%%%%%%%%%%%%%%%%%%%%%%%%%%%%%%%%%%%%%%%%%%%%%%%%%%%%%%%

\bibliographystyle{IEEEtran}
\bibliography{references}

\begin{thebibliography}{10}
\providecommand{\url}[1]{#1}
\csname url@rmstyle\endcsname
\providecommand{\newblock}{\relax}
\providecommand{\bibinfo}[2]{#2}
\providecommand\BIBentrySTDinterwordspacing{\spaceskip=0pt\relax}
\providecommand\BIBentryALTinterwordstretchfactor{4}
\providecommand\BIBentryALTinterwordspacing{\spaceskip=\fontdimen2\font plus
\BIBentryALTinterwordstretchfactor\fontdimen3\font minus
  \fontdimen4\font\relax}
\providecommand\BIBforeignlanguage[2]{{%
\expandafter\ifx\csname l@#1\endcsname\relax
\typeout{** WARNING: IEEEtran.bst: No hyphenation pattern has been}%
\typeout{** loaded for the language `#1'. Using the pattern for}%
\typeout{** the default language instead.}%
\else
\language=\csname l@#1\endcsname
\fi
#2}}

\bibitem{elmeseiry2021detailed}
N.~Elmeseiry, N.~Alshaer, and T.~Ismail, ``A detailed survey and future
  directions of unmanned aerial vehicles ({UAVs}) with potential
  applications,'' \emph{Aerospace}, vol.~8, no.~12, p. 363, 2021.

\bibitem{rashad2020fully}
R.~Rashad, J.~Goerres, R.~Aarts, J.~B. Engelen, and S.~Stramigioli, ``Fully
  actuated multirotor {UAV}s: A literature review,'' \emph{IEEE Robotics \&
  Automation Magazine}, vol.~27, no.~3, pp. 97--107, 2020.

\bibitem{michieletto2018fundamental}
G.~Michieletto, M.~Ryll, and A.~Franchi, ``Fundamental actuation properties of
  multirotors: Force--moment decoupling and fail--safe robustness,'' \emph{IEEE
  Trans. on Robotics}, vol.~34, no.~3, pp. 702--715, 2018.

\bibitem{tadokoro2017maneuverability}
Y.~Tadokoro, T.~Ibuki, and M.~Sampei, ``Maneuverability analysis of a
  fully-actuated hexrotor uav considering tilt angles and arrangement of
  rotors,'' \emph{IFAC-PapersOnLine}, vol.~50, no.~1, pp. 8981--8986, 2017.

\bibitem{michieletto2017control}
G.~Michieletto, M.~Ryll, and A.~Franchi, ``Control of statically hoverable
  multi-rotor aerial vehicles and application to rotor-failure robustness for
  hexarotors,'' in \emph{Int. Conf. on Robotics and Automation}.\hskip 1em plus
  0.5em minus 0.4em\relax IEEE, 2017, pp. 2747--2752.

\bibitem{michieletto2017nonlinear}
G.~Michieletto, A.~Cenedese, L.~Zaccarian, and A.~Franchi, ``Nonlinear control
  of multi-rotor aerial vehicles based on the zero-moment direction,''
  \emph{IFAC-PapersOnLine}, vol.~50, no.~1, pp. 13\,144--13\,149, 2017.

\bibitem{michieletto2020hierarchical}
------, ``Hierarchical nonlinear control for multi-rotor asymptotic
  stabilization based on zero-moment direction,'' \emph{Automatica}, vol. 117,
  p. 108991, 2020.

\bibitem{rashad2019port}
R.~Rashad, F.~Califano, and S.~Stramigioli, ``Port-hamiltonian passivity-based
  control on {SE} (3) of a fully actuated uav for aerial physical interaction
  near-hovering,'' \emph{IEEE Robotics and Automation Letters}, vol.~4, no.~4,
  pp. 4378--4385, 2019.

\bibitem{antonello2018dual}
A.~Antonello, G.~Michieletto, R.~Antonello, and A.~Cenedese, ``A dual
  quaternion feedback linearized approach for maneuver regulation of rigid
  bodies,'' \emph{IEEE Control Systems Letters}, vol.~2, no.~3, pp. 327--332,
  2018.

\bibitem{arizaga2019adaptive}
J.~M. Arizaga, H.~Casta{\~n}eda, and P.~Castillo, ``Adaptive control for a
  tilted-motors hexacopter {UAS} flying on a perturbed environment,'' in
  \emph{Int. Conf. on Unmanned Aircraft Systems}.\hskip 1em plus 0.5em minus
  0.4em\relax IEEE, 2019, pp. 171--177.

\bibitem{flores2022robust}
G.~Flores, A.~M. de~Oca, and A.~Flores, ``Robust nonlinear control for the
  fully actuated hexa-rotor: Theory and experiments,'' \emph{IEEE Control
  Systems Letters}, vol.~7, pp. 277--282, 2022.

\bibitem{rajappa2015modeling}
S.~Rajappa, M.~Ryll, H.~H. B{\"u}lthoff, and A.~Franchi, ``Modeling, control
  and design optimization for a fully-actuated hexarotor aerial vehicle with
  tilted propellers,'' in \emph{Int. Conf. on Robotics and Automation}.\hskip
  1em plus 0.5em minus 0.4em\relax IEEE, 2015, pp. 4006--4013.

\bibitem{franchi2018full}
A.~Franchi, R.~Carli, D.~Bicego, and M.~Ryll, ``Full-pose tracking control for
  aerial robotic systems with laterally bounded input force,'' \emph{IEEE
  Trans. on Robotics}, vol.~34, no.~2, pp. 534--541, 2018.

\bibitem{hamandi2023full}
M.~Hamandi, I.~Al-Ali, L.~Seneviratne, A.~Franchi, and Y.~Zweiri, ``Full-pose
  trajectory tracking of overactuated multi-rotor aerial vehicles with limited
  actuation abilities,'' \emph{IEEE Robotics and Automation Letters}, 2023.

\end{thebibliography}

\end{document}